\documentclass[
floatfix,
reprint,
superscriptaddress,
amsmath,amssymb,
aps, pra,
]{revtex4-2}

\usepackage{amsmath,amssymb,amsfonts}
\usepackage{algorithm,algpseudocode}
\usepackage{graphicx}
\usepackage{booktabs}
\usepackage{nicematrix}
\usepackage{svg}
\usepackage{textcomp}
\usepackage{xcolor}
\usepackage{hyperref}
\usepackage{physics}
\usepackage{diagbox}
\usepackage{bm}
\usepackage{array}
\usepackage{slashbox}
\usepackage{graphicx}
\usepackage{dcolumn}
\usepackage{ragged2e}
\usepackage{placeins}

\usepackage{tikz}
\usetikzlibrary{decorations.pathreplacing}
\usetikzlibrary{quantikz2, arrows.meta, positioning, calc}

\newcommand{\BigO}{\mathcal{O}}

\newcommand{\ceil}[1]{\lceil #1\rceil}

\DeclareMathOperator{\cvar}{CVaR}

\begin{document}

\title{CVaR-Assisted Custom Penalty Function for Constrained Optimization}

\author{Xin Wei LEE}
\email{xwlee@smu.edu.sg}

\author{Hoong Chuin LAU}
\email{hclau@smu.edu.sg (Corresponding Author)}
\affiliation{School of Computing and Information Systems, Singapore Management University}


\begin{abstract}
We propose a slack-free penalty formulation for constrained binary optimization that eliminates auxiliary slack variables and preserves the feasibility structure of the original problem. The proposed approach introduces a nonlinear custom penalty function to enforce inequality constraints directly in the objective function. To address the computational challenges associated with evaluating nonlinear penalties in variational quantum algorithms, we employ the finite-sampling method that avoids the exponential complexity required by exact expectation computation. Furthermore, we integrate the Conditional Value-at-Risk (CVaR) objective to improve optimization robustness and guide the search toward high-quality solutions.
The proposed framework is evaluated on instances of the multi-dimensional knapsack problem, a classical benchmark in combinatorial optimization.
We showcase that the proposed custom-penalty formulation combined with CVaR sampling achieves improved optimality gaps and more consistent performance compared with conventional slack-based QUBO formulations.
We also tested our method on the Quantinuum H2 quantum hardware, demonstrating the possibilities of training small-sized VQE on noisy trapped-ion quantum computers.
The results suggest that careful penalty design can play a critical role in enabling quantum and hybrid quantum-classical algorithms for constrained optimization problems that arise in operations research.
\end{abstract}

\maketitle

\section{Introduction}
Variational quantum algorithms have become an emerging field of study to tackle problems that are difficult to be solved in classical computers,
including the notorious combinatorial optimization problems (COP), which are usually NP-hard.
The quantum approximate optimization algorithm (QAOA)~\cite{farhi2014quantum} has been widely studied due to its relation to Trotterized adiabatic quantum computation and quantum 
annealing~\cite{farhi:qaa,farhi2019quantum}, and hence its richness in mathematical properties.
On the other hand, the variational quantum eigensolver (VQE)~\cite{vqe} is a more general approach, which is built upon a higher expressivity, problem-agnostic ansatz to offer a higher
degree of freedom to the trial state.
Previous work has shown that VQE usually requires a shallower quantum circuit to achieve the same performance as QAOA~\cite{vqe-vs-qaoa}.
This leads to an optimization landscape that is less likely to have barren plateaus and less noise when running on real quantum hardware.

Solving unconstrained COP with VQE is straightforward. The binary variables are mapped into the Pauli-Z operator, which then sums into the Hamiltonian.
This is done in such a way that the objectives corresponding to each combination of the binary variables are encoded in the eigenvalues of the Hamiltonian. 
A trial state $\ket{\psi(\bm\theta)}$ is evolved with a parameterized quantum circuit,
then the expectation of the Hamiltonian $\expval{H}{\psi(\bm\theta)}$ is then minimized or maximized to approximate the solution of the original problem.
The Max-cut problem is a benchmark problem in the quantum computing community to evaluate the performances of VQE and QAOA.
However, constrained COP is more complicated. The loss function is usually added with penalty terms to enforce constraints.
The end result of this formulation is usually a quadratic unconstrained binary optimization (QUBO) program.
Others use feasibility-preserving circuits to search for solutions only in the feasible space, and these frameworks are categorized under a more general
Quantum Alternating Operator Ansatz~\cite{new-qaoa,grover-mixer,th-gm-qaoa,constraint-preserving-qaoa,prog-mis}.
Recently, a feasibility-guiding loss function has been proposed along with its ansatz that requires multi-controlled Toffoli gates~\cite{toffoli-vqa}.
The drawback for the feasibility-preserving circuits is clear---the oracle circuit used to create a superposition of feasibility states is usually deep and requires extra ancilla qubits,
which results in more noise in quantum hardware.
For the multi-controlled Toffoli approach, although the best known implementation of an $n$-qubit Toffoli gate only requires $6n+2$ controlled-NOT (CNOT) gates with depth $\BigO(\log n)$,
there is still a considerable amount of CNOT gates, and the error rate of two-qubit gates in quantum hardware is roughly 10 times of one-qubit gates.
In this work, we focus on the discussion of penalized QUBO, which has several drawbacks, such as the unfaithful optimization landscape that does not reflect the original problem, 
and the introduction of slack variables to the problem that complicates the search space even further.
The slack variables are dummy variables that are introduced in the penalty terms of the inequality constraints.

The focus naturally shifted to the elimination of slack variables in the formulation of problems with inequality constraints.
Some works have proposed the possibility of adopting the Lagrangian formulation, which does not require the addition of slack variables, to be solved using quantum
annealers~\cite{ohzeki-inequality,ohzeki-subgradient} and also in gate-based VQE~\cite{vqec,dual-vqe}.
Although in~\cite{ohzeki-inequality} a step function is embedded within the partition function to create an effective Hamiltonian to be sampled, it is entirely different in nature from the method proposed in
gate-based models.
Moreover, the Lagrangian loss function is updated using the subgradient method.

On the other hand, the unbalanced penalty approach is also proposed as an effort to approximate the exponential penalty function with a second-order polynomial,
which makes the penalized loss function native to quantum annealers~\cite{ub-penalty,ub-dwave}.
The work in ~\cite{ohzeki-inequality} hinted at the use of a Heaviside step function to replace the quadratic function of the inequality constraints, which can be conveniently achieved by sampling
a transformed Hamiltonian on a quantum annealer. 
For gate-based quantum circuits, a recent work that proposed IF-QAOA~\cite{if-qaoa,efficient-qaoa} formulates a loss function embedded with a step function,
with the QAOA evolution of the stepped Hamiltonian implemented by a quantum phase estimation (QPE) subroutine 
to calculate the values of inequality constraints and an oracle that marks the sign of the value given by QPE.
However, the QPE subroutine requires many additional gates and ancillary qubits when implemented on near-term noisy quantum devices,
which appears to be conceptually misaligned with QAOA that is suited for near-term devices.

In our previous preliminary work~\cite{step-qce}, we proposed a method to directly evaluate the expectation of a step-embedded Hamiltonian,
with the variational state evolved by a problem-agnostic hardware-efficient ansatz (HEA)~\cite{hea}.
We showed that our step penalty achieves higher feasibility rate and optimality rate than the unbalanced penalty method for the Multiple Knapsack problem, and will not be repeated in this paper. A limitation of that work is that it requires an exponential overhead of $\BigO(2^t)$, where $t$ is the number of variables in a single inequality constraint. This imposes a limitation on applying the method to tackle more complex constraint optimization problems, such as the Multi-Dimensional Knapsack problem (MDKP), which is an example of the class of strongly NP-hard problems that are hard to compute even on classical machines \cite{mdkp-overview,challenges-qo}.
Motivated to improve the previous method, in this work, we propose the use of the finite sampling (FS) method for the expectation calculation, which eliminates the exponential overhead.
This is similar to the filtering VQE (F-VQE) proposed in~\cite{f-vqe}, except that the F-VQE applies the filtering operators on the quantum state,
where we applied them on the constraint Hamiltonians.
Our method can be further enhanced by seamlessly combining with some of the optimization tricks, e.g. the Conditional Value-at-Risk (CVaR), to produce a better optimality gap. 
We test our method on MDKP and our results include an evaluation on noiseless simulations, as well as the Quantinuum H2 noisy quantum hardware. 
Both show that the method is a more promising approach than the conventional slack-QUBO formulation.

In summary, the main contribution of this paper is to diverge from the literature by utilizing a problem-agnostic HEA; and by evaluating a nonlinear Heaviside step-embedded Hamiltonian directly through
finite sampling enhanced with CVaR optimization, we eliminate the need for extra ancilla qubits, high CNOT gates and deep oracle circuits, thereby maintaining a hardware-friendly footprint while 
rigorously enforcing inequality constraints in combinatorial optimization problems. Furthermore, our approach preserves the true problem cost structure exactly on the feasible region, which matters when 
the reported objective value must be interpretable as the true problem cost rather than a penalized approximation (such as \cite{ohzeki-subgradient}). 

\section{Background}
In this section, we introduce the preliminaries of our previous work, as well as the finite sampling (FS) and the Conditional Value-at-Risk (CVaR) method 
used in this work. We also introduce the multi-dimensional knapsack problem, which is used as a benchmark problem in this work.

\begin{figure}
    \centering
    \includegraphics[width=0.95\linewidth]{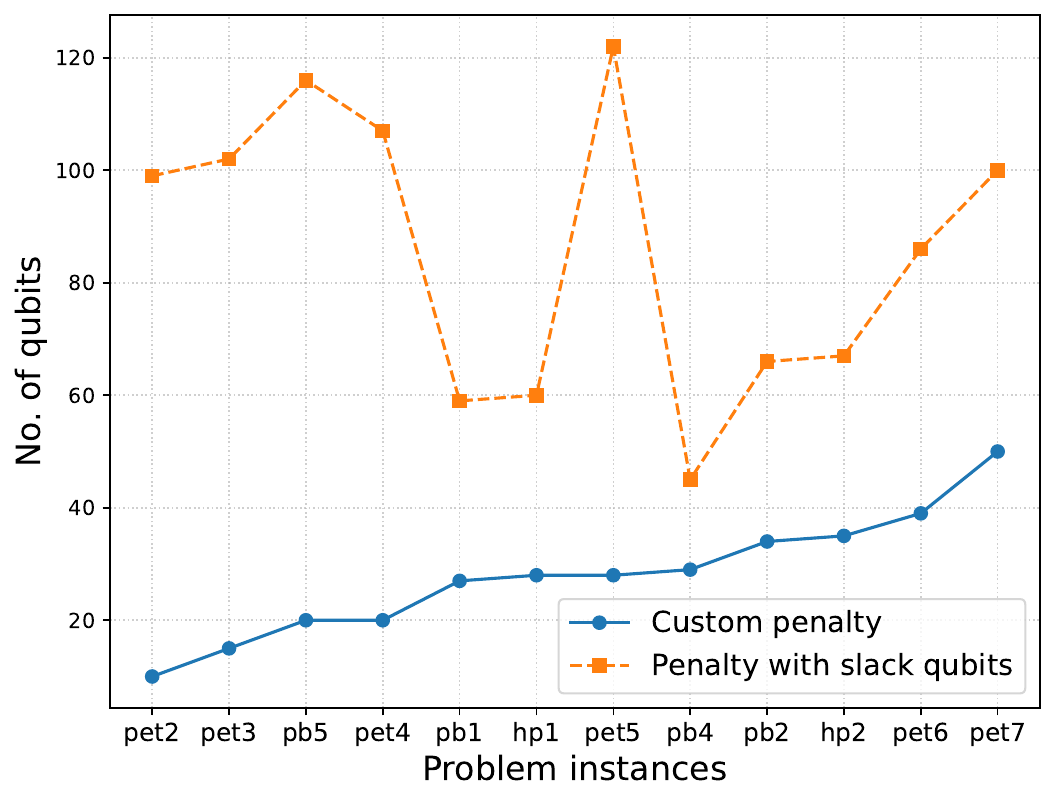}
    \caption{Comparison between the number of qubits required for the custom penalty and the common slack variable formulation. The problem instances are the multi-dimensional knapsack problem used in this work.}
    \label{fig:num-qubits-comparison}
\end{figure}

\subsection{Custom Penalty}
Consider a constrained binary optimization problem
\begin{align}
    \min_x\ & f(x) \\ 
    \text{s.t.}\ & g_i(x) = 0,\quad i = 1,...,m, \\
    & h_j(x) \leq 0,\quad j= 1,...,k,
\end{align}
where each bit in $x\in\{0,1\}^n$ represents a binary variable of the problem.
A standard formulation of QUBO enforces each constraint by assigning penalty factors (or penalty coefficients) to penalty functions that give non-zero values when the solution $x$ violates a constraint,
causing the value of the QUBO loss function to be high when constraints are violated during the minimization. 
Equality constraints are enforced by introducing the term $\lambda g(x)^2$, where $\lambda$ is the penalty factor. 
Inequality constraints, on the other hand, are trickier.
The inequalities are converted to equalities by introducing the slack (dummy) variables $y_l$, then enforced in the loss function by squaring the
whole term. Hence, the resulting QUBO loss function for the constrained problem is written as 
\begin{multline}
    \mathcal{L}_{\text{slack}}(x) =
    f(x) + \sum_{i=1}^m \lambda_0^{(i)} g_i(x)^2 \\
    + \sum_{j=1}^k \lambda_1^{(j)}\left(h_j(x) + \sum_{l=0}^{N-1}2^ly_l \right)^2.
    \label{eqn:loss-slack}
\end{multline}
There are a few issues in the above formulation: 1) there is no clear line dividing the feasible solutions and the infeasible solutions for the inequality penalty term.
Although there exist certain values of $y$ that will cause the penalty term to be zero when the constraint is satisfied (i.e. $x$ being a feasible solution), there are also other values of $y$ that make
the penalty term large even when the constraint is satisfied. This causes the possible solutions to be flooded with redundant combinations of $y$ that are in fact not required to be searched during 
the optimization. This further deteriorates the optimization landscape of variational quantum algorithms that are already full of local minima.
2) When Eq.~(\ref{eqn:loss-slack}) is converted into a Hamiltonian in terms of Pauli operators to be solved in a quantum computer, the slack variables must be represented in binary form (hence $2^ly_l$).
This requires $\lceil\log_2 s_j\rceil$ extra variables to be added to the problem, where $s_j$ is the largest possible value for $h_j(x)$. 
This can add a considerable amount of extra qubits when the magnitude of $s_j$ is large, which yields larger search space, a common phenomenon in variational algorithms---the barren plateaus,
and more noise on real quantum hardware.
In Fig.~\ref{fig:num-qubits-comparison}, we show a comparison of the number of qubits needed for our custom penalty method and the common slack variable formulation.
The instances are the multi-dimensional knapsack instances used as the target problems in this work.
It shows that the system can add up to an extra of 80 qubits with the slack variables included, which is an enormous amount considering the hardware available nowadays (about 100 qubits).

A new method is proposed in~\cite{step-qce} that includes a custom penalty function $\xi(\cdot)$ in the loss function to enforce the inequality constraints:
\begin{equation}
    \mathcal{L}_{\text{custom}}(x) = f(x) + \sum_{j=1}^k\lambda_j\ \xi[h_j(x)].
    \label{eqn:custom-loss}
\end{equation}
$\xi(\cdot)$ is usually nonlinear, and must be able to divide the feasible solutions ($h_j(x)\le 0$) and the infeasible solutions ($h_j(x)>0$) with sufficiently large difference in its values.
One such candidate is the Heaviside step function: 
\begin{equation}
    \Theta[h_j(x)] = 
    \begin{dcases}
        1 & h_j(x) > 0 \\
        0 & h_j(x) \le 0 
    \end{dcases}
    \label{eqn:step-func}
\end{equation}
The value of the penalty function is then amplified by the penalty factor $\lambda$. The penalty factor must be sufficiently large to offset the largest possible value of the objective $f(x)$.
Eq.~(\ref{eqn:custom-loss}) is converted to a Hamiltonian to be solved using a quantum computer:
\begin{equation}
    H = H_f + \sum_{j=1}^k \lambda_j\ \xi(H_j).
\end{equation}
$H_f$ is the Hamiltonian corresponds to the objective $f(x)$ and $H_j$'s are those that correspond to the constraints $h_j(x)$. The Hamiltonians are expressed in terms of one-local and two-local Pauil-Z's.
The ground energy of $H$ is then found using variational methods to obtain the solution that minimizes (\ref{eqn:custom-loss}).
This is equivalent to minimizing the expectation $\expval{H}{\psi(\bm\theta)}$, where $\ket{\psi(\bm\theta)}$ is a trial state prepared by the variational ansatz circuit.
$\expval{H}$ can be broken down into the sum of the expectation of its components:
\begin{equation}
    \expval{H} = \expval{H_f} + \sum_{j=1}^k \lambda_j\expval{\xi(H_j)}.
    \label{eqn:custom-ham}
\end{equation}
Since $H_f$ is a sum of Pauli-Z's, we can further break it down into the expectation of $Z$'s, i.e., $\expval{H_f} = \sum_{i,j} \expval{Z_iZ_j} + \sum_i\expval{Z_i}$. 
Each $\expval{Z}$ or $\expval{ZZ}$ corresponds to the measurements in the $Z$ basis on different qubits in the variational ansatz. 
However, since $\xi(\cdot)$ is a nonlinear function, we cannot further break down $\expval{\xi(H_i)}$ in terms of $\expval{Z}$ or $\expval{ZZ}$, and it has to be calculated by other means.

We derived in~\cite{step-qce} that it requires a time complexity of $\BigO(2^t)$ to perform a full diagonalization on $\expval{\xi(H_j)}$,
where $t$ is the number of distinct variables in the constraint $h_j(x)$.
Although the custom penalty method has a great degree of freedom in choosing the penalty function, computing its expectation will not be easy on large problems. 
This calls for a method to improve the computational cost of the custom penalty.

\subsection{Finite Sampling (FS) and CVaR}\label{sec:fs-cvar}
The expectation in Eq.~(\ref{eqn:custom-ham}) can also be calculated with the results obtained by sampling the circuit. In simulations or real quantum devices, we construct a quantum circuit that evolves into the 
final state $\ket{\psi}$, then $\ket{\psi}$ is sampled a finite number of times (where we call \emph{shots}).
We call this the finite sampling (FS).
Each shot returns a solution bit-string $x(\bm\theta)$ that corresponds to a computational basis state of $\ket{\psi(\bm\theta)}$.
We use the notation $\hat{E}$ to represent this estimator and $E$ to represent the exact expectation of $H$:
\begin{equation}
    \hat{E}(\bm\theta) \approx \expval{H}{\psi(\bm\theta)} = E(\bm\theta).
\end{equation}
The estimator $\hat{E}(\bm\theta)$ is then given as
\begin{equation}
    \hat{E}(\bm\theta) = \sum_{i=0}^{2^n-1} \mathcal{L}(x(\bm\theta)=i)p_i.
    \label{eqn:exp}
\end{equation}
$i$ represents the binary eigenstate $x$ in the integer form, $\mathcal{L}(x(\bm\theta)=i)$ is the loss function evaluated at $i$.
$p_i$ is the probability of obtaining the eigenstate $i$, calculated by $p_i=m_i/M$, where $m_i$ is the number of shots of $i$ sampled and $M$ is the total number of shots.
By manipulating Eq.~(\ref{eqn:exp}), $\hat{E}$ can be calculated by an alternative method:
\begin{equation}
    \hat{E}(\bm\theta) = \frac{1}{M}\sum_{m=1}^M \mathcal{L}(x^{(m)}(\bm\theta)),
    \label{eqn:finite-sampling}
\end{equation}
where $x^{(m)}$ is the solution sampled from the $m$-th shot and $\mathcal{L}$ is the loss function.
Eq.~(\ref{eqn:finite-sampling}) shows that $\hat{E}$ is in fact the mean of the sampled eigenvalues.
The expectation can now be estimated with a fixed budget $M$ for the Hamiltonian $H$ of any form, without needing to populate all the eigenvalues with $\BigO(2^t)$.
Note that to obtain the full information of the state $\ket{\psi}$, $M$ usually scales exponentially with the number of qubits.
Here, we assume that only partial information of the state is required for optimization, which we will connect to the method that is going to be discussed next.

The Conditional Value-at-Risk (CVaR) method is introduced in~\cite{cvar} to sample diagonal Hamiltonians in a more efficient way.
Similar to the scenario above, we take $M$ samples from the quantum circuit, each denoted by $x^{(m)}$. The samples are then sorted as $x^{(1)},x^{(2)},\ldots,x^{(M)}$,
such that their loss values are in non-decreasing order,
$\mathcal{L}(x^{(1)})\le \mathcal{L}(x^{(2)})\le \ldots \le \mathcal{L}(x^{(M)})$. Here, we still assume a minimization problem, i.e., we want to find the lowest value in $\mathcal{L}$.
A confidence level $0<\alpha<1$ is chosen, then the expectation is calculated by including only the $\lceil\alpha M\rceil$ samples with the lowest eigenvalues ($\lceil\cdot\rceil$ is the ceiling function):
\begin{equation}
    \cvar(\alpha;\bm\theta) = \frac{1}{\ceil{\alpha M}} \sum_{m=1}^{\ceil{\alpha M}} \mathcal{L}(x^{(m)}(\bm\theta)).
    \label{eqn:cvar}
\end{equation}
The loss function $\expval{H}$ is replaced by $\cvar(\alpha;\bm\theta)$. It is important to note that when $\alpha=1$, the CVaR computes the original expectation $\expval{H}$ by including all the samples.
It is claimed that minimizing CVaR leads to a smoother optimization landscape, compared to the landscape of $\expval{H}$.
Moreover, the landscape of CVaR is more resilient to noise by filtering out the higher eigenvalues of the Hamiltonian.

\subsection{Multi-dimensional Knapsack}
The multi-dimensional knapsack problem (MDKP) is a generalized version of the 0-1 knapsack problem to multiple constraints. Each constraint may represent a limit in a certain ``dimension'', e.g.,
weight, volume, number of items, etc. The MDKP is formally defined as
\begin{alignat}{2}
    \max_x & \sum_{i=1}^n v_ix_i  \label{eqn:mdkp-start} \\
    \text{s.t.} & \sum_{i=1}^n w_{ij}x_i \leq W_j,& \quad j = 1,...,d \\
    & x_i\in\{0,1\}, & i=1,...,n, \label{eqn:mdkp-end}
\end{alignat}
where \\
\begin{tabular}{lll}
    $x_i$ & : & Decision variable of the item $i$. \\
    $v_i$ & : & Value/profit of the item $i$. \\
    $w_{ij}$ & : & Weight of item $i$ in dimension $j$. \\
    $W_j$ & : & Capacity (limit) of the weight in dimension $j$.
\end{tabular} \\

There is only one knapsack in this problem setting. The decision variable $x_i$ decides whether the item $i$ is included in the knapsack (1 if included, 0 otherwise).
Each $j$ is a constraint on a dimension, with a total of $d$ constraints.
Even for $d=2$, there is no known Efficient Polynomial-Time Approximation Scheme (EPTAS) to approximate the solution of MDKP~\cite{eptas-knapsack},
which means that MDKP is believed to be much harder than the standard 0-1 knapsack.

\begin{table}
    \centering
    \caption{The multi-dimensional knapsack instances, their number of variables and their number of dimensions used in the simulations.}
    \begin{NiceTabularX}{0.8\linewidth}{cXc}
        \toprule
        \toprule
        \textbf{Instance} & \textbf{\# variables} & \textbf{\# dimensions} \\
        \midrule
        \textbf{hp1} & 28 & 4 \\
        \textbf{hp2} & 35 & 4 \\
        \textbf{pb1} & 27 & 4 \\
        \textbf{pb2} & 34 & 4 \\
        \textbf{pb4} & 29 & 2 \\
        \textbf{pb5} & 20 & 10 \\
        \textbf{pet2} & 10 & 10 \\
        \textbf{pet3} & 15 & 10 \\
        \textbf{pet4} & 20 & 10 \\
        \textbf{pet5} & 28 & 10 \\
        \textbf{pet6} & 39 & 5 \\
        \textbf{pet7} & 50 & 5 \\
        \bottomrule
        \bottomrule
    \end{NiceTabularX}    
    \label{tab:instaces}
\end{table}

\section{Methodology}\label{sec:method}
We first discuss our framework in general.
The objective is clear: we want to minimize the expectation stated in Eq.~(\ref{eqn:custom-ham}) with the standard VQE routine, i.e., a quantum circuit to generate the trial state and a classical optimizer
to minimize the value of $\expval{H}$.
A problem-agnostic ansatz is used to generate the trial state $\ket{\psi(\bm\theta)}$. Otherwise, it would require an extra effort to design a Hamiltonian-dependent ansatz if the custom penalty
Hamiltonian were to be raised to the exponent, such as the circuit proposed in~\cite{if-qaoa}.
Then, we obtain $M$ samples of the trial state $\ket{\psi(\bm\theta)}$.
Each sampled solution $x^{(i)}$ is substituted back into the original custom loss function $\mathcal{L}_\text{custom}$ to obtain its corresponding objective (eigenvalue of the Hamiltonian $H$).
This is then summed up using Eq.~(\ref{eqn:finite-sampling}) to obtain the expectation $\expval{H}$ of the given $M$ samples. 
This sampling step can also be replaced with (\ref{eqn:cvar}) to compute the CVaR of $H$ for a pre-selected confidence level $\alpha$. 

In our simulations, we solve the MDKP stated in Eq.~(\ref{eqn:mdkp-start})--(\ref{eqn:mdkp-end}). First, the custom loss function $\mathcal{L}_\text{custom}(x)$ is formed by summing up the objective
$f(x)$ with the custom penalty terms $\xi[h_j(x)]$. We use the Heaviside step function $\Theta(\cdot)$ [Eq.~(\ref{eqn:step-func})] as the penalty function. 
This loss function can then be written as
\begin{equation}
    \mathcal{L}_\text{MDKP}(x) = -\sum_{i=1}^n v_ix_i + \sum_{j=1}^d \sum_{i=1}^n \lambda_j\Theta\left(\sum_{i=1}^n w_{ij}x_i - W_j\right).
    \label{eqn:mdkp-loss}
\end{equation}
The objective is negated to convert it to a minimization problem.
The problem instances solved are several instances selected from the OR-Library~\cite{orlib} and used as benchmark instances in~\cite{monit-benchmark},
ranging from 10 variables to 50 variables, while the dimension of MDKP ranges from 2 to 10. 
The details of each instance (a total of 12 instances) are stated in Table~\ref{tab:instaces}.
For the penalty factor $\lambda_j$, we use the upper bound method and set a global penalty $\lambda=\lambda_j=2\lambda_\text{UB},\ \forall j$, where
\begin{equation}
    \lambda_\text{UB} = \sum_{i=1}^n v_i,
\end{equation}
so that one violation of the constraint will offset the objective in~(\ref{eqn:mdkp-start}).

We choose a single layer HEA with $R_Y$ parameterized rotational gates and CZ entangling gates between adjacent qubits.
The ansatz is initialized with a layer of parameterized $R_Y$.
The trial state generated $\ket{\psi(\bm\theta)}$ is formalized as
\begin{equation}
    \ket{\psi(\bm\theta)} = \bigotimes_{i=1}^n R_Y(\theta_{n+i}) \bigotimes_{i=1}^{n-1} CZ_{i,i+1} \bigotimes_{i=1}^n R_Y(\theta_i),
    \label{eqn:ansatz}
\end{equation}
where $\bm\theta = (\theta_1,\theta_2,\ldots,\theta_{2n})$. $R_Y(\theta_i)$ denotes the first layer of $R_Y$ gates acting on qubit $i$ with the parameter $\theta_i$.
$R_Y(\theta_{n+i})$ denotes the second layer of $R_Y$ gates acting on qubit $i$ with the parameter $\theta_{n+i}$.
$CZ_{i,j}$ denotes the unparameterized controlled-Z gate acting on the control qubit $i$ and target qubit $j$.
We justify our choice of a single layer ansatz by assuming that a single layer is sufficient to find the solution of a COP, which is a computational basis state (an eigenstate of a diagonal Hamiltonian).
Therefore, we also assume that introducing more layers will complicate the optimization search space, leading to more local minima 
and barren plateaus~\cite{barren_vqa2021,Cerezo_2021,bp-qnn,bp-review} in the optimization landscape.

\begin{figure*}[t]
    \centering
    \includegraphics[width=\linewidth]{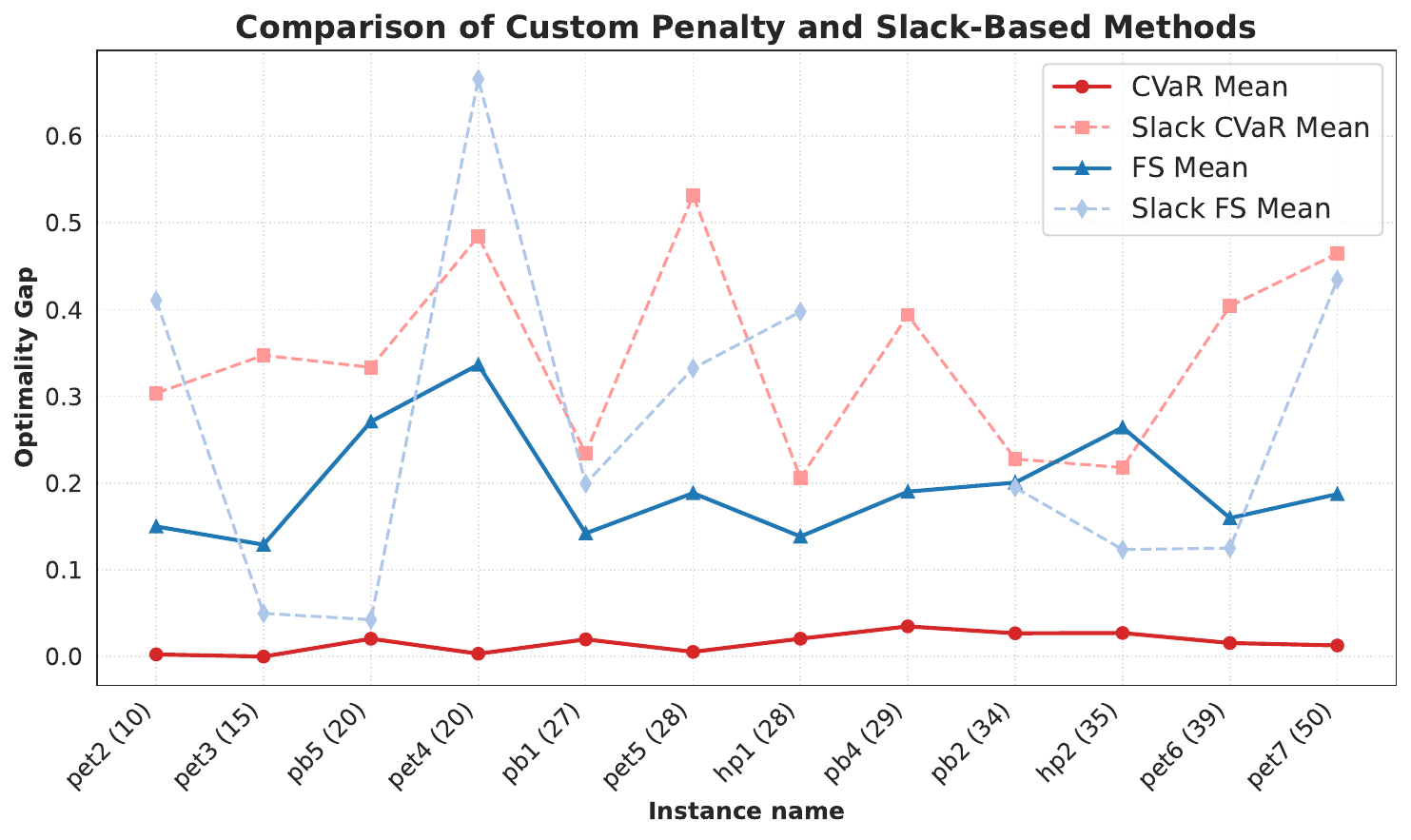}
    \caption{Comparison between the mean optimality gaps between slack and custom penalty (non-slack) formulations. The horizontal axis is labeled with the name of each instance along with the number of qubits
    in the bracket. The solid lines are the mean optimality gaps for the custom penalty method. The dashed lines are the mean optimality gaps for the formulation with slack variables. The blue and red lines
    shows the mean optimality gaps for the finite sampling (FS) and Conditional Value-at-Risk (CVaR) respectively. The mean for custom penalty is taken by averaging the optimality gap for 20 trials
    (different parameter initialization).}
    \label{fig:slack-vs-noslack}
\end{figure*}

\section{Analysis on the Sampling Cost}
It is mentioned in Sec.~\ref{sec:fs-cvar} that the computation of the expectation by FS (and CVaR) can be estimated with a fixed budget (number of samples) $M$,
instead of the $\BigO(2^t)$ complexity to compute the eigenvalues of the constrained Hamiltonian.
This estimation comes with a sampling error $\epsilon$, which is governed by Hoeffding's inequality:
\begin{equation}
    \mathrm{Pr}(|\hat{E}(\bm\theta)-E(\bm\theta)|\geq \epsilon) \leq 2\exp \left(-\frac{2M\epsilon^2}{R^2} \right),
    \label{eqn:hoeffding}
\end{equation}
where $R=b-a$ is the range of the loss function $\mathcal{L}(x)\in[a,b]$.
Consider taking $M$ samples of a random variable with mean $E$, the sampled mean $\hat{E}$ is different from the mean $E$ by some sampling error $\epsilon$.
Hoeffding's inequality states that the probability of the absolute difference between $\hat{E}$ and $E$ is greater than or equal to $\epsilon$ (LHS of (\ref{eqn:hoeffding})) is upper bounded by some value 
that decays exponentially with $M$, $\epsilon^2$, and $1/R^2$ (RHS of (\ref{eqn:hoeffding})).
Rearranging (\ref{eqn:hoeffding}) gives us the number of shots $M$ required to achieve the sampling error $\epsilon$ with confidence $1-\delta$:
\begin{equation}
    M\geq \frac{R^2}{2\epsilon^2}\ln\left(\frac{2}{\delta}\right),
    \label{eqn:shot-ineq}
\end{equation}
where $\delta=\mathrm{Pr}(|\hat{E}(\bm\theta)-E(\bm\theta)|\geq \epsilon)$. 
For most real world COPs or the MDKP instances considered in this work, $E$ is usually a large integer, so $\epsilon$ can also be large for rough estimation, e.g. $\epsilon\approx100$ to $1000$.
The range of the loss function $R$ depends on the formulation and the problem instance.
For the formulation in (\ref{eqn:mdkp-loss}), the lowest value for the loss $\mathcal{L}_\text{MDKP}(x)$ is the ground truth of the MDKP instance
$-C_\text{true}$; the highest value is the sum of the penalty factors, which in our case is $\lambda d$, where $d$ is the dimension of MDKP. The range $R$ in our case is hence
\begin{equation}
    R = C_\text{true} + 2d\sum_{i=1}^n v_i.
\end{equation}
To keep the number of shots required $M$ low, it is important to balance $R$ and $\epsilon$ so that $R^2/\epsilon^2$ is low. 
If we want the error $\epsilon$ to be small, then $R$ also has to be small. 
This is where rescaling the problem coefficients becomes important, as it is possible to rescale the problem such that it reduces the range $R$ of the loss function, leading to less number of shots required.
Several works have discussed the effect of rescaling problem coefficients having positive effects in optimization and helping convergence~\cite{pt-weighted,align-qaoa,ccop-qite}.
However, appropriate rescaling is required to avoid loss of information in the original problem objectives~\cite{ccop-qite}.

For CVaR, the reasoning is similar. Let $p_{x^*}$ be the probability of sampling the quasi-optimum $x^*$ in the converged state $\ket{\psi(\bm\theta^*)}$.
The probability of sampling $x^*$ using CVaR is
\begin{equation}
    p_\alpha = 
    \begin{dcases}
        \frac{p_{x^*}}{\alpha} & \text{if }\ p_{x^*}<\alpha \\
        1 & \text{if }\ p_{x^*}\geq \alpha.
    \end{dcases}
\end{equation}
This is because CVaR creates a \emph{tail} that concentrates all the lowest eigenvalues by sorting them in ascending order.
Therefore, if the proportion of the tail to the entire sample exceeds $\alpha$, i.e., $p_{x^*} = M_{x^*}/M\geq \alpha$, then the entire tail consists of $x^*$.
Otherwise, if $p_{x^*}<\alpha$, then $p_\alpha = M_{x^*}/\lceil\alpha M\rceil \approx p_{x^*}/\alpha$, as only $\lceil\alpha M\rceil$ samples are taken but $M_{x^*}$ stays the same.
With that in mind, we can deduce that if FS needs $M$ samples to achieve an estimator $\hat{E}$, then CVaR only needs $M_\alpha = \lceil\alpha M\rceil$ samples to achieve $\hat{E}'\leq \hat{E}$
which always holds due to the sorting mechanism. Combining with (\ref{eqn:shot-ineq}), we obtain
\begin{equation}
    M_\alpha \geq \alpha M_\text{FS} \geq \frac{\alpha R^2}{2\epsilon^2}\ln\left(\frac{2}{\delta}\right).
\end{equation}
Hence, we get the minimum number of shots required by CVaR as
\begin{equation}
    M_\alpha \geq \frac{\alpha R^2}{2\epsilon^2}\ln\left(\frac{2}{\delta}\right).
\end{equation}
This serves as a baseline for determining the number of shots for CVaR. The number of shots for CVaR is $1/\alpha$-fold less compared to the number of shots for FS.
However, note that CVaR requires sorting the eigenvalues in every function evaluation, which takes another $\BigO(kn\log n)$ for algorithms such as merge sort and heap sort, for the entire optimization process.
$k$ is the number of function evaluations and usually scales with the number of parameters. 

\begin{figure*}
    \centering
    \includegraphics[width=\linewidth]{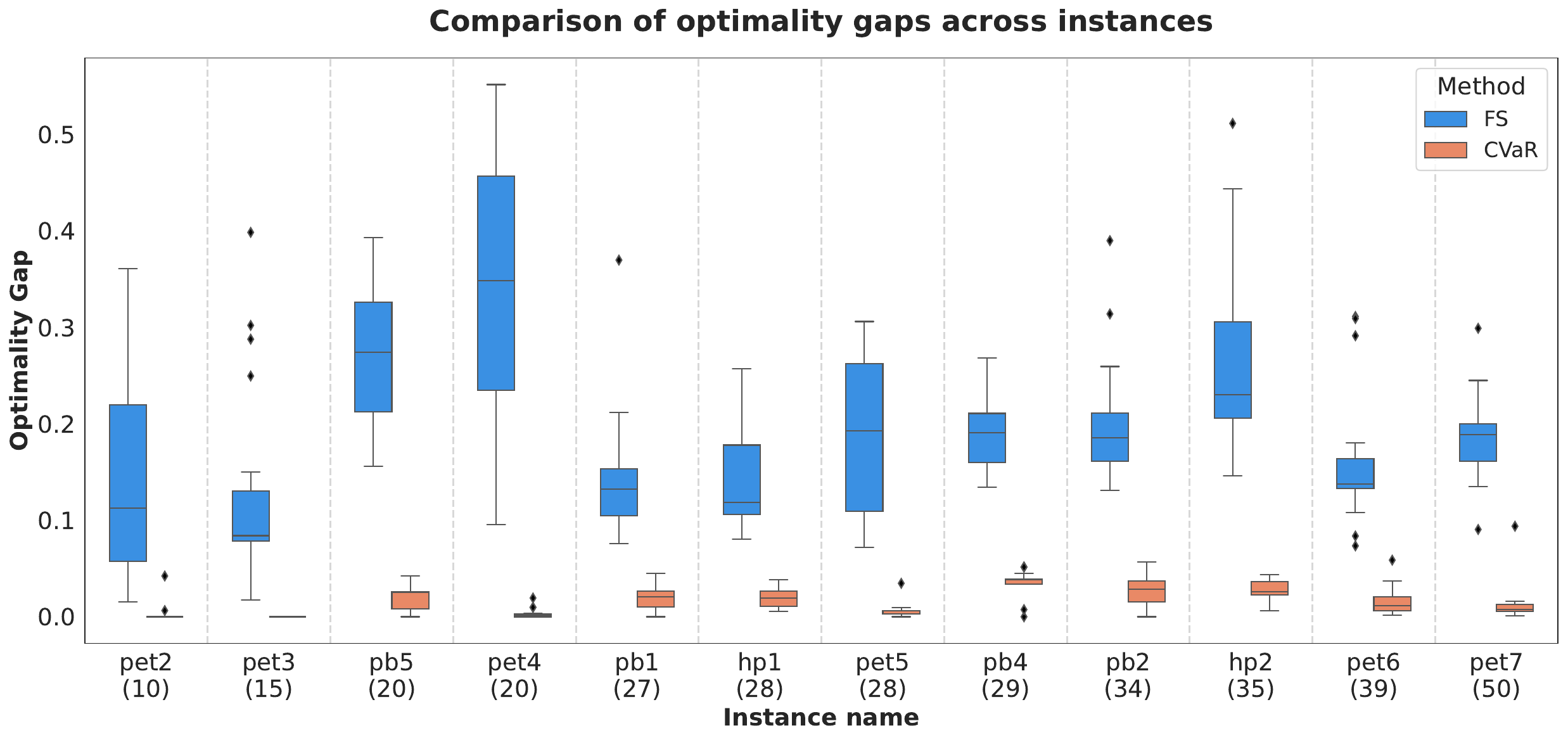}
    \caption{Comparison between the optimality gaps for different methods: finite sampling (FS) and Conditional Value-at-Risk (CVaR). The horizontal axis is labeled with the name of each instance along with
    the number of qubits in the bracket. The boxes shows the optimality gap of each instance with 20 different trials (different parameter initializations). The blue boxes show the optimality gap obtained
    with finite sampling (FS). The orange boxes show the optimality gap obtained with Conditional Value-at-Risk (CVaR).}
    \label{fig:mdkp-optgap}
\end{figure*}

\begin{figure*}
    \centering
    \includegraphics[width=\linewidth]{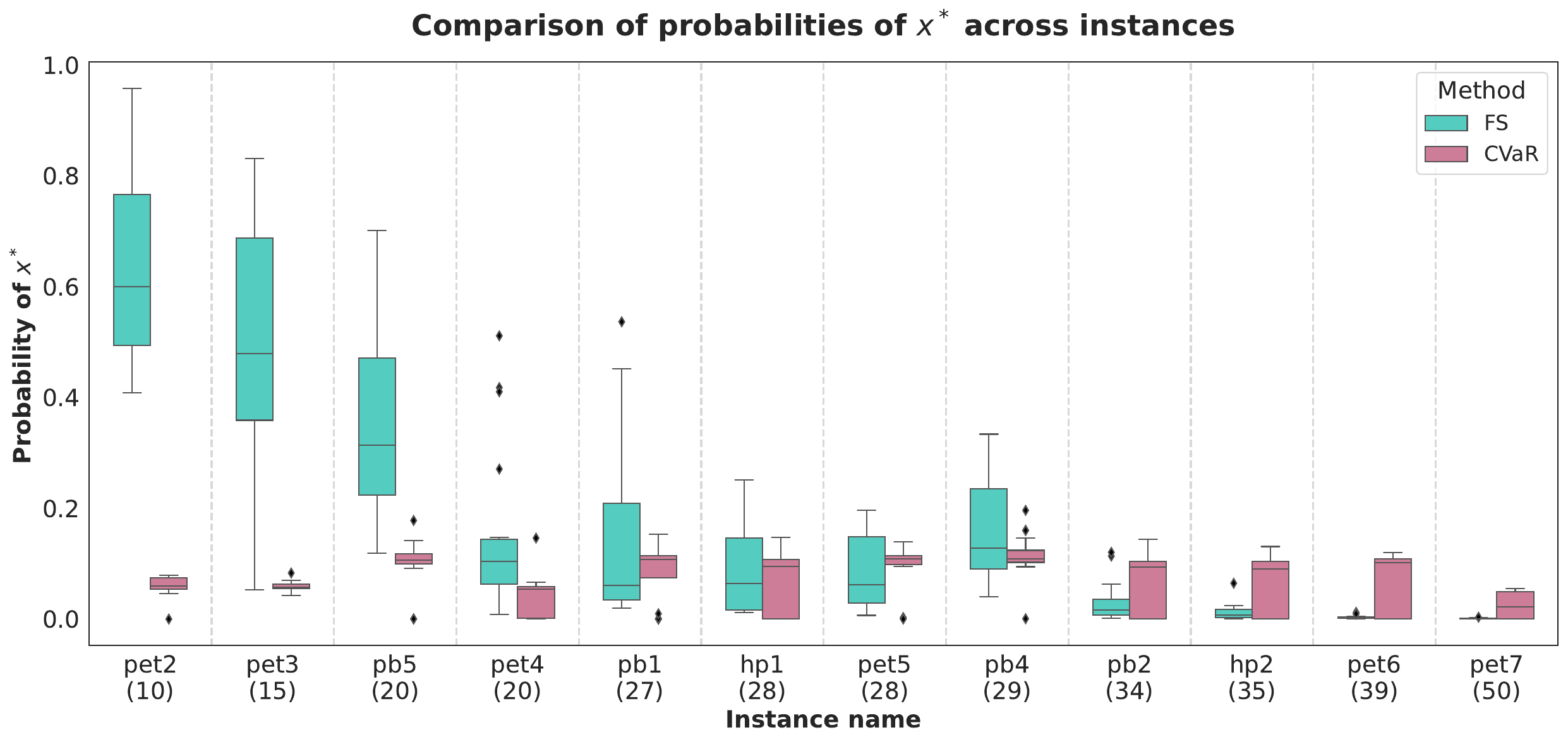}
    \caption{Comparison between the probability of sampling $x^*$ (quasi-optimum) for different methods: finite sampling (FS) and Conditional Value-at-Risk (CVaR).
    The horizontal axis is labeled with the name of each instance along with
    the number of qubits in the bracket. The boxes shows the optimality gap of each instance with 20 different trials (different parameter initializations). The turquoise boxes show the optimality gap obtained
    with finite sampling (FS). The pink boxes show the optimality gap obtained with Conditional Value-at-Risk (CVaR).}
    \label{fig:mdkp-opt-prob}
\end{figure*}


\section{Results and Discussions}
\subsection{Noiseless simulation}\label{sec:noiseless-sim}
The details of the simulation are stated as follows:
\begin{enumerate}
    \item \textbf{Problem instances}: MDKP benchmark instances that are selected from the OR-Library~\cite{orlib}.
    The number of variables and the number of dimensions (constraints are specified in Table~\ref{tab:instaces}).
    \item \textbf{Simulator}: IBM Qiskit Aer Matrix Product State Simulator.
    \item \textbf{Ansatz}: HEA with an initial $R_Y$ layer, a linearly entangled CZ chain, and then another $R_Y$ layer, as stated in Eq.~(\ref{eqn:ansatz}).
    \item \textbf{Sampling}: Each circuit is sampled with 4000 shots. For CVaR, the confidence level is set as $\alpha=0.1$.
    \item \textbf{Initialization}: Optimization is initialized with 20 sets of different random parameters.
    \item \textbf{Optimizer}: POWELL optimizer from Qiskit Algorithms with \verb+maxfev=10000+, \verb+xtol=1e-4+ and other hyperparameters left default.
    \item \textbf{Evaluation metric}: Optimality gap, probability of quasi-optima, and number of function evaluations.
\end{enumerate}

The optimality gap $\Delta$ is defined as
\begin{equation}
    \Delta = 1 - \frac{C_\text{VQE}}{C_\text{true}}.
    \label{eqn:opt-gap}
\end{equation}
The optimality gap denotes how close the solution objective given by VQE is to that of the original ground truth, with 0 being exactly the ground truth,
and larger values mean further from the ground truth.
The solution objective $C_\text{VQE}$ is obtained by substituting the quasi-optimal solution $x^*$ back into the objective in Eq.~(\ref{eqn:mdkp-start}), i.e., $C_\text{VQE} = \mathcal{L}(x^*)$.
The quasi-optimal solution $x^*$ is defined as the eigenstate of $H$ (bit-string of computational basis) that has the highest probability in the sampled distribution of $\ket{\psi(\bm\theta^*)}$,
where $\bm\theta^*$ are the quasi-optimal parameters obtained from VQE.
For CVaR, since the solutions are already sorted in $\mathcal{L}(x^{(1)})\le \mathcal{L}(x^{(2)})\le \ldots \le \mathcal{L}(x^{(M)})$, we treat $x^{*(1)}$ as the quasi-optimal solution instead of $x^*$. 
Note that the asterisk denotes the quasi-optimal solution obtained from the convergence of VQE. In this case, $C_\text{VQE} = \mathcal{L}(x^{*(1)})$.
In practice, we observe only a subtle difference between $x^{*(1)}$ and $x^*$ when using CVaR.

We compare our custom penalty method with the commonly used slack-based formulation. The custom penalty is implemented with two different sampling methods:
finite sampling (FS) and Conditional Value-at-Risk (CVaR)~\cite{cvar}. The optimality gaps for custom penalty and slack-based formulation, each with different sampling methods,
are compared in Fig.~\ref{fig:slack-vs-noslack}.
First, we look at the optimality gap $\Delta$---which represents the quality of the solution and directly tells us which method solves the MDKP with the least amount of error.
The optimality gaps for the slack-based formulation are taken from the prior benchmarking work in~\cite{monit-benchmark}, while the optimality gaps for the custom penalty are done using simulation settings
stated in Sec.~\ref{sec:method}.
For fair comparison, the number of shots for each circuit run and the optimizer used, including its hyperparameters, are tallied with those in~\cite{monit-benchmark}.
In Fig.~\ref{fig:slack-vs-noslack}, the $\Delta$ for each MDKP instance are plotted. The instances in the horizontal axis are sorted with the number of qubits used in the custom penalty formulation,
i.e., the number of variables in the objective of the original problem.
The main observation here is that the custom penalty generally has lower mean $\Delta$ than the slack-based method under the same simulation settings.
The mean $\Delta$ for custom penalty is taken by averaging the optimality gaps for 20 trials (different parameter initializations).
For FS, 7 out of 12 instances (pet2, pet4, pb1, pet5, hp1, pb4, pet7) has lower mean $\Delta$ for the custom penalty method.
The instance pb4 has failed to reach a feasible solution with the slack formulation.
Other instances are able to achieve feasibility for both the slack-based and the custom penalty methods.
For CVaR, we can see that the custom penalty for all instances have lower $\Delta$ than the slack-based method, with a low optimality gap $0<\Delta<0.1$ that is very close to the solution.
Although not shown in the figure, all the instances achieved feasibility with the custom penalty formulation. 
This shows the effectiveness and advantage of using custom penalty compared to adding slack qubits into the formulation.

\begin{figure*}
    \centering
    \includegraphics[width=\linewidth]{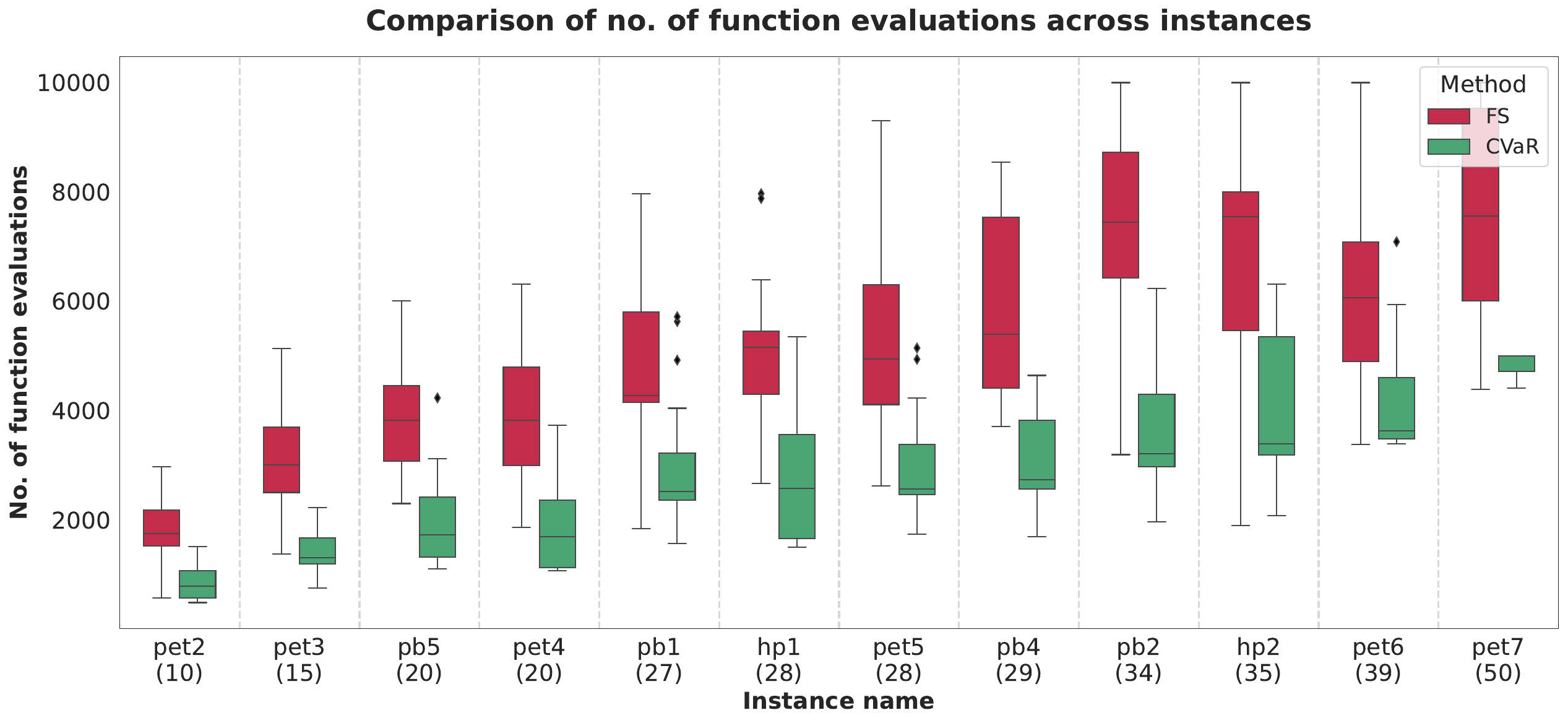}
    \caption{Comparison between the number of function evaluations for different methods: finite sampling (FS) and Conditional Value-at-Risk (CVaR).
    The horizontal axis is labeled with the name of each instance along with
    the number of qubits in the bracket. The boxes shows the optimality gap of each instance with 20 different trials (different parameter initializations). The red boxes show the optimality gap obtained
    with finite sampling (FS). The green boxes show the optimality gap obtained with Conditional Value-at-Risk (CVaR).}
    \label{fig:mdkp-nfev}
\end{figure*}

Fig.~\ref{fig:mdkp-optgap} shows the boxplot for the optimality gap of each instance with the custom penalty formulation. We compare two different sampling methods: FS and CVaR.
The horizontal axis labels the instances sorted by their respective number of qubits required. Interestingly, the result does not show any clear degradation in the optimality gap as the number of qubits 
in the system increases. For example, for FS, the median gap for the largest instance (pet7) is comparable to or even lower than that of smaller instances (pet4).
This suggests that the inherent structure or difficulty of specific instances plays a more significant role in optimization success than the dimensionality alone. 
Mid-range instances like pet4 and hp2 show a wider vertical spread (interquartile range), indicating higher sensitivity to initialization or local minima for these configurations.
The side-by-side comparison between FS (blue boxes) and CVaR (orange boxes) shows the optimality gap of FS is further enhanced by combining CVaR with custom penalty.
In all instances, CVaR achieves lower median optimality gaps and lower variances (smaller boxes) compared to FS.
The reduced height of the CVaR boxes across several larger instances demonstrates that the CVaR objective provides better solution consistency and higher reliability.
While both methods encounter occasional outliers (represented by the circles above the whiskers), the CVaR approach consistently pushes the lower bound of the gap closer to zero,
signifying its effectiveness in navigating the optimization landscapes of these MDKP instances.

Fig.~\ref{fig:mdkp-opt-prob} shows the probability of sampling the quasi-optimal solution $p_{x^*}$ for each instance.
For FS, the median $p_{x^*}$ decreases with the number of qubits, showing the \emph{curse of dimensionality}~\cite{bp-review}.
This is caused by an exponential increase in the number of possible states in the quantum state space, hence the number of shots also needs to increase exponentially to observe a particular state.
Since the number of shots is fixed in this simulation, this is an expected phenomenon.
However, this effect is mitigated by applying CVaR. We can see that the median probability of $x^*$ for CVaR does not decrease with the number of qubits and concentrates at around 0.1.
This is because we have selected the confidence level of CVaR to be $\alpha=0.1$, so only top 10\% of the eigenvalues will be used to evaluate the expectation.
In other words, in the sampled distribution for CVaR, the probability of $x^*$ only needs to be 10\% (or 0.1) to achieve the maximum expectation and hence converging the optimization. 

Fig.~\ref{fig:mdkp-nfev} shows a comparison in the number of function evaluations performed by the optimizer between FS and CVaR. 
A general trend shows that the number of evaluations increases with the number of qubits.
It is also observed that the CVaR usually takes less number of evaluations than FS.


\subsection{Hardware results}
Our custom penalty method with both FS and CVaR is also tested on the Quantinuum H2 series trapped ion quantum computers. 
The specifications and details of Quantinuum H2 can be found in its product data sheet~\cite{quantinuum-h2-data}.
We use the same HEA as specified in (\ref{eqn:ansatz}) with an adaptation to the H2 native gate set.
The native gates used are the $R_Z$ gate, the PhasedX gate $\phi_x(\gamma,\beta)$ and the ZZPhase gate $\phi_{zz}(\beta)$:
\begin{align}
    \phi_X(\gamma,\beta) & = R_Z(\beta)R_X(\gamma)R_Z(-\beta) \\
    \phi_{ZZ}(\beta) & = e^{-i\beta(Z\otimes Z)},
\end{align}
where $R_X$ and $R_Z$ represent the typical Pauli $X$ and $Z$ rotations respectively. 
Therefore, our HEA is implemented as 
\begin{align}
    R_Y(\gamma) & = \phi_X\left(\gamma,\frac{\pi}{2}\right) \\
    CZ_{i,j} & = e^{-i\frac{\pi}{4}} \phi_{ZZ}\left(\frac{\pi}{4}\right) R_{Z_i}\left(\frac{\pi}{4}\right) R_{Z_j}\left(\frac{\pi}{4}\right).
\end{align}
The global phase of $e^{-i\pi/4}$ is omitted during the actual implementation of the circuit as it does not affect the result of the end-circuit measurements.
It is worth noting that the gate functions in the Pytket circuit simulation package
for Quantinuum devices take in angles that are normalized with $\pi$, so the angles $\gamma$ and $\beta$ will be multiplied by $\pi$ in the Pytket quantum circuit.

\begin{figure}
    \centering
    \includegraphics[width=0.95\linewidth]{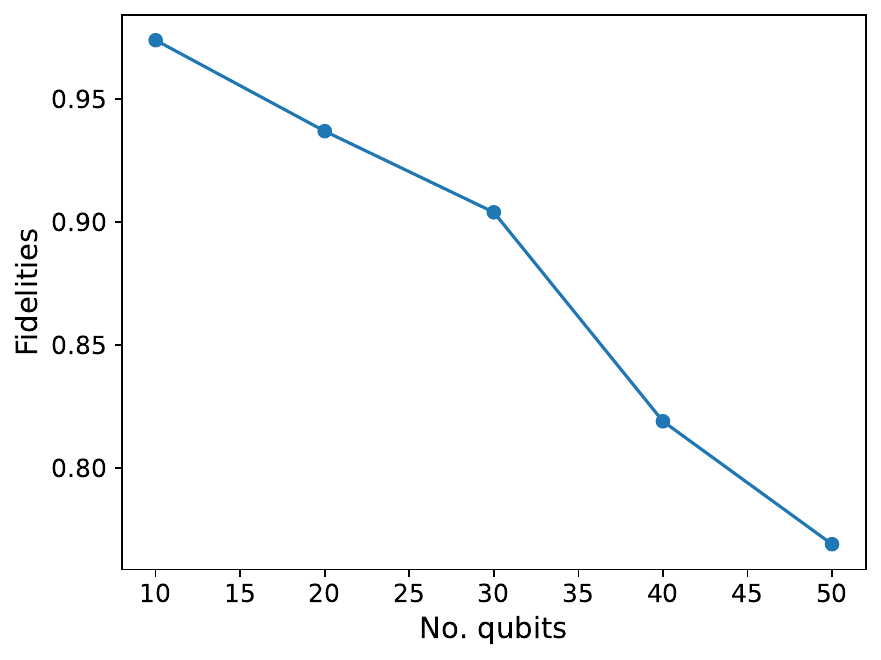}
    \caption{Fidelities of our HEA circuit versus the number of qubits on Quantinuum H2.
    The circuit $U^\dagger U$ is run on the quantum hardware and the probability of measuring the $\ketbra{0}^{\otimes n}$ state (for $n$ qubits) is recorded as the fidelity,
    where $U$ denotes our HEA circuit.}
    \label{fig:quantinuum-benchmark}
\end{figure}

\begin{table*}
    \centering
    \caption{Comparison of the optimality gaps for different simulation backends in solving the MDKP instances.
    For noiseless simulation, we use the Matrix Product State (MPS) simulator provided in Qiskit.
    For the noisy simulation and real quantum hardware, we use the Quantinuum H2 noisy emulator to train the VQE and both the emulator and the real hardware to test the optimal circuit 
    (the circuit instance with the optimal parameters). Same set of initial parameters are used to train the VQE for the noiseless and noisy emulators.
    The values show the test results of a run of the optimal circuit on the noiseless MPS, the H2 emulator, and the H2 quantum hardware. 
    The objective for each instance is recorded with its optimality gap calculated in the parentheses. }
    \begin{NiceTabularX}{0.9\linewidth}{ccXXXc}
    \toprule
    \toprule
        \Block{2-1}{\textbf{Instance}} & \Block{2-1}{\textbf{Method}} & \Block{1-4}{\textbf{Objective (optimality gap)}} & & & \\
        \cmidrule{3-6}
        & & \textbf{Noiseless} & \textbf{Noisy emulator} & \textbf{Hardware} & \textbf{Optimal} \\
        \cmidrule{3-6}
        \Block{2-1}{pet2} & FS & 83369 (0.042) & 83183 (0.045) & 83183 (0.045) & 87061 \\
        \cmidrule{2-6}
        & CVaR & 87061 (0.0) & 87061 (0.0) & 87061 (0.0) & 87061 \\
    \bottomrule
    \bottomrule
    \end{NiceTabularX}
    \label{tab:hardware-results}
\end{table*}

\begin{figure}
    \centering
    \includegraphics[width=0.95\linewidth]{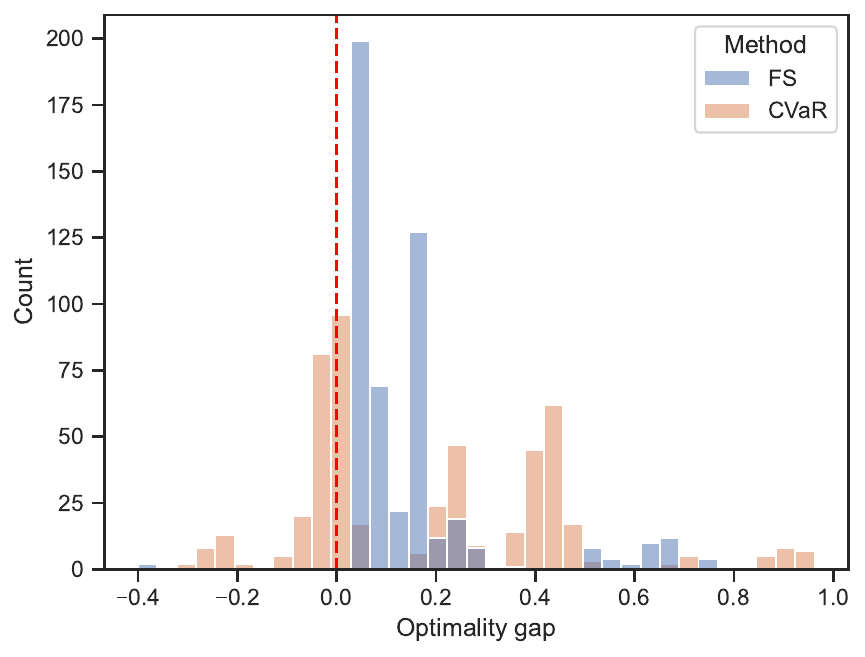}
    \caption{Distribution of the optimality gaps of the sampled bit-string for the pet2 instance. The counts record the number of times each bit-string with its corresponding optimality
    gap is sampled, with a total of 500 shots. Different bit-strings with the same optimality gap are summed together as a bar. The red dashed line represents an optimality gap of 0, 
    which corresponds to the optimal solution. The negative values represent infeasible solutions that has higher objective than the optimal objective.}
    \label{fig:pet2-counts}
\end{figure}

Fig.~\ref{fig:quantinuum-benchmark} shows the fidelity test we ran on the H2 quantum hardware.
The test is conducted by appending the $n$-qubit single-layer HEA $U$ with its Hermitian $U^\dagger$ and measuring its fidelity,
i.e., the probability of measuring the $\ketbra{0}^{\otimes n}$ state.
1000 samples are taken from the circuit and the probability of measuring the $\ketbra{0}^{\otimes n}$ state is recorded as the fidelity. 
The result shows that the fidelity of the circuit decreases almost linearly as the number of qubits increases.
Although the hardware can achieve a fidelity of 0.97 at 10 qubits, it gradually decreases to 0.77 at 50 qubits.
This shows that the training of VQE might still be promising at a small number of qubits.

We use the same optimizer settings as stated in Sec.~\ref{sec:noiseless-sim} for the H2 runs.
However, due to the limitation of the quantum resources that we were able to access,
we used the H2 emulator to train the VQE and the H2 quantum hardware only to test the optimal angles. 
The final solution is obtained by running the optimal circuit (the circuit instance with the optimal angles) on the H2 quantum hardware.
Also, we only managed to run the smallest MDKP instances: pet2 (10 qubits) with one random initialization due to the limitation in the hardware resource.
We also set the number of shots to 500 in this case.
Nevertheless, we were able to obtain decent results for the pet2 instance. 

Fig.~\ref{fig:pet2-counts} shows the distribution of the optimality gaps of the sampled bit-strings from the optimal circuit for pet2. 
Different bit-strings with the same optimality gap are summed together as a bar. 
Recall that the optimality gap of 0 corresponds to the optimal solution, and values further than zero represent solutions further from the optimum.
Note that the negative values represent solutions that are infeasible but have higher objectives than the optimum. 
It can be observed that FS has almost no shots in the infeasible regime (only one at $\Delta=-0.4$) due to the enforcement of the custom penalty,
but has failed to achieve the optimal solution.
On the other hand, CVaR has varied solutions in the spectrum, as it only considers the solutions of the best $\ceil{\alpha M}$ shots and ignores other solutions when computing its loss,
but was able to find the optimal solution in roughly 100 shots within 500 shots.

Table~\ref{tab:hardware-results} records the optimality gap of different backends testing the optimal circuit obtained from the VQE trained on the H2 emulator.
For noiseless simulation, we use the Matrix Product State (MPS) simulator provided in Qiskit. For the noisy simulation and real quantum
hardware, we use the Quantinuum H2 emulator to train the VQE and both the emulator and the real hardware to test the
optimal circuit. The same set of initial parameters are used to train the VQE for noiseless and noisy emulators.
For FS, both noisy emulation and hardware yield the same optimality gap of 0.045, which is close to the 0.045 given by noiseless simulation.
For CVaR, all backends have achieved the optimal solution, showing promising results in VQE training on noisy quantum backends.

\section{Conclusion and Future Works}
This work serves as a contribution on previous work that proposed a custom penalty to solve constrained optimization problems with VQE.
Our previous method \cite{step-qce} is limiting in that computing the expectation for the nonlinear custom penalty function requires the knowledge
of at most $2^t$ eigenvalues of the constraint Hamiltonian, causing the implementation to have a time complexity of $\BigO(2^t)$.

In this work, we replace the expectation computation component in VQE with sampling strategies, namely finite sampling (FS) and Conditional Value-at-Risk (CVaR),
so that the expectation can be computed with a fixed number of shots.
The estimator of the expectation is then computed as the sampled mean of the loss values for the bit-strings sampled from the final quantum state. This opens the computational pathway to tackle practical constraint optimization problems such as the Multi-dimensional knapsack problem (MDKP) which we could not do in our previous work due to computational overheads.  
Inherently, this imposes a sampling error when estimating the expectation, which scales with the number of shots and the range of the loss function.
We derived the number of shots required for the simulation with the Hoeffding's bound, in terms of the range of the values for the loss function.
We test our method on the MDKP benchmark instances obtained from ORLib.
The problems range from 10 to 50 qubits (variables), with dimensions of 4 to 10. 

We compare our custom penalty loss function with the typical QUBO formulation with slack variables.
The results show that the custom penalty loss function has lower optimality gaps (closer to the optimal solution) in 7 out of 12 instances that we considered,
due to fewer qubits used and hence less complicated optimization landscape.
With CVaR, all the instances show lower optimality gaps with custom penalty compared to the slack penalty.
Next, we compare FS and CVaR with the custom penalty formulation with the following metrics: optimality gap, number of function evaluations, and the probability of sampling the quasi-optimal solution $p_{x^*}$.
We observe that although $p_{x^*}$ for FS decreases exponentially with the problem size (due to the curse of dimensionality), the optimality gap does not explicitly decrease with the problem size.
Even with the largest 50-qubit problem (pet7) that we consider, the median optimality gap for FS is around 0.2. For all metrics, we can see significant improvement by adopting a CVaR loss function.
For pet7, the optimality gap can reach a median of 0.0076 ($10^{-3}$ scale) for CVaR.

A direct empirical comparison against smoothed-penalty formulations, such as the unbalanced penalization method ~\cite{ub-penalty,ub-dwave}, on the same MDKP benchmark set and other hard combinatorial optimization problems would further demonstrate the efficacy of our approach. We leave this comparative study, along with an ablation on ansatz depth and penalty factor sensitivity, for future work.

In terms of hardware implementation, we tested the pet2 instance on the Quantinuum H2 quantum hardware. The variational circuit is trained on a noisy H2 emulator and the optimal circuit is tested on the
real hardware. The noisy VQE was able to achieve results that are very close to the noiseless simulation ($\approx 0.3\%$ difference with respect to the optimal objective) 
for FS, and was able to find the optimal solution with CVaR.
Despite the shortcomings of the currently available hardware, the results are promising in training VQE on a noisy backend for small number of qubits.

From the practical point of view, the limitations of training a VQE on currently available quantum hardware are apparent.
This includes long queue times and lost connections to the server in the midst of optimization, which might cause variational quantum optimization to be impractical.
Recent advances in fault-tolerant quantum computing have suggested a paradigm shift towards non-variational algorithms for quantum optimization,
e.g. Grover's adaptive search~\cite{gas}, non-variational Quantum Alternating Operator Ansatz (QAOA)~\cite{nv-qaoa}, Quantum Imaginary Time Evolution (QITE)~\cite{qite-seminal}, etc.,
along with the utilization of non-unitary operations like qubit reset and mid-circuit measurements~\cite{qubit-reset}. 

\section*{Data and Code Availability}
The data that support the findings of this article are publicly available at \url{https://github.com/SMU-Quantum/custom-penalty}.
The implementation of the discussed core method can be installed with 
\verb+pip install custom-penalty+. (The links will be made public upon the acceptance of the manuscript.)

\section*{Funding Declaration}
This project is supported by the National Research Foundation, Singapore through the National Quantum Office, hosted in A*STAR, under its Quantum Engineering Programme 3.0 Funding 
Initiative (W24Q3D0002) and its Advanced Quantum Computing Infrastructure Funding Initiative (S25Q8D9001).

\bibliography{ccop,qaoaref,qiteref}

@article{ub-penalty,
   title={Unbalanced penalization: a new approach to encode inequality constraints of combinatorial problems for quantum optimization algorithms},
   volume={9},
   ISSN={2058-9565},
   url={http://dx.doi.org/10.1088/2058-9565/ad35e4},
   DOI={10.1088/2058-9565/ad35e4},
   number={2},
   journal={Quantum Science and Technology},
   publisher={IOP Publishing},
   author={Montañez-Barrera, J A and Willsch, Dennis and Maldonado-Romo, A and Michielsen, Kristel},
   year={2024},
   month=apr, pages={025022}
}

@misc{vqec,
      title={Variational Quantum Eigensolver with Constraints (VQEC): Solving Constrained Optimization Problems via VQE}, 
      author={Thinh Viet Le and Vassilis Kekatos},
      year={2024},
      eprint={2311.08502},
      archivePrefix={arXiv},
      primaryClass={quant-ph},
      url={https://arxiv.org/abs/2311.08502}, 
}

@inproceedings{ub-dwave,
   title={Improving Performance in Combinatorial Optimization Problems with Inequality Constraints: An Evaluation of the Unbalanced Penalization Method on D-Wave Advantage},
   url={http://dx.doi.org/10.1109/QCE57702.2023.00067},
   DOI={10.1109/qce57702.2023.00067},
   booktitle={2023 IEEE International Conference on Quantum Computing and Engineering (QCE)},
   publisher={IEEE},
   author={Montañez-Barrera, J. A. and van den Heuvel, Pim and Willsch, Dennis and Michielsen, Kristel},
   year={2023},
   month=sep, pages={535–542}
}

@misc{ohzeki-inequality,
    title={Solving Inequality-Constrained Binary Optimization Problems on Quantum Annealer}, 
    author={Kouki Yonaga and Masamichi J. Miyama and Masayuki Ohzeki},
    year={2020},
    eprint={2012.06119},
    archivePrefix={arXiv},
    primaryClass={quant-ph},
    url={https://arxiv.org/abs/2012.06119}, 
}

@article{ohzeki-subgradient,
   title={Subgradient Method Using Quantum Annealing for Inequality-Constrained Binary Optimization Problems},
   volume={94},
   ISSN={1347-4073},
   url={http://dx.doi.org/10.7566/JPSJ.94.054003},
   DOI={10.7566/jpsj.94.054003},
   number={5},
   journal={Journal of the Physical Society of Japan},
   publisher={Physical Society of Japan},
   author={Takabayashi, Taisei and Goto, Takeru and Ohzeki, Masayuki},
   year={2025},
   month=may
}

@INPROCEEDINGS{step-qce,
  author={Lee, Xin Wei and Lau, Hoong Chuin},
  booktitle={2025 IEEE International Conference on Quantum Computing and Engineering (QCE)}, 
  title={Implementing Slack-Free Custom Penalty Function for QUBO on Gate-Based Quantum Computers}, 
  year={2025},
  volume={01},
  number={},
  pages={2112-2119},
  doi={10.1109/QCE65121.2025.00231}
}

@article{cvar,
   title={Improving Variational Quantum Optimization using CVaR},
   volume={4},
   ISSN={2521-327X},
   url={http://dx.doi.org/10.22331/q-2020-04-20-256},
   DOI={10.22331/q-2020-04-20-256},
   journal={Quantum},
   publisher={Verein zur Forderung des Open Access Publizierens in den Quantenwissenschaften},
   author={Barkoutsos, Panagiotis Kl. and Nannicini, Giacomo and Robert, Anton and Tavernelli, Ivano and Woerner, Stefan},
   year={2020},
   month=apr, pages={256}
}

@article{eptas-knapsack,
    author = {Kulik, Ariel and Shachnai, Hadas},
    title = {There is no EPTAS for two-dimensional knapsack},
    year = {2010},
    issue_date = {July, 2010},
    publisher = {Elsevier North-Holland, Inc.},
    address = {USA},
    volume = {110},
    number = {16},
    issn = {0020-0190},
    url = {https://doi.org/10.1016/j.ipl.2010.05.031},
    doi = {10.1016/j.ipl.2010.05.031},
    journal = {Inf. Process. Lett.},
    month = jul,
    pages = {707–710},
    numpages = {4}
}

@misc{orlib,
    author = {J.E. Beasly},
    title = {OR-Library},
    howpublished = {\url{https://people.brunel.ac.uk/~mastjjb/jeb/orlib/mknapinfo.html}},
    year = {2018}
}

@misc{monit-benchmark,
      title={A Comparative Study of Quantum Optimization Techniques for Solving Combinatorial Optimization Benchmark Problems}, 
      author={Monit Sharma and Hoong Chuin Lau},
      year={2025},
      eprint={2503.12121},
      archivePrefix={arXiv},
      primaryClass={quant-ph},
      url={https://arxiv.org/abs/2503.12121}, 
}

@article{mdkp-overview,
    title = {The multidimensional 0–1 knapsack problem: An overview},
    journal = {European Journal of Operational Research},
    volume = {155},
    number = {1},
    pages = {1-21},
    year = {2004},
    issn = {0377-2217},
    doi = {https://doi.org/10.1016/S0377-2217(03)00274-1},
    url = {https://www.sciencedirect.com/science/article/pii/S0377221703002741},
    author = {Arnaud Fréville}
}

@article{challenges-qo,
   title={Challenges and opportunities in quantum optimization},
   volume={6},
   ISSN={2522-5820},
   url={http://dx.doi.org/10.1038/s42254-024-00770-9},
   DOI={10.1038/s42254-024-00770-9},
   number={12},
   journal={Nature Reviews Physics},
   publisher={Springer Science and Business Media LLC},
   author={Abbas, Amira and Ambainis, Andris and Augustino, Brandon and Bärtschi, Andreas and Buhrman, Harry and Coffrin, Carleton and Cortiana, Giorgio and Dunjko, Vedran and Egger, Daniel J. and Elmegreen, Bruce G. and Franco, Nicola and Fratini, Filippo and Fuller, Bryce and Gacon, Julien and Gonciulea, Constantin and Gribling, Sander and Gupta, Swati and Hadfield, Stuart and Heese, Raoul and Kircher, Gerhard and Kleinert, Thomas and Koch, Thorsten and Korpas, Georgios and Lenk, Steve and Marecek, Jakub and Markov, Vanio and Mazzola, Guglielmo and Mensa, Stefano and Mohseni, Naeimeh and Nannicini, Giacomo and O’Meara, Corey and Tapia, Elena Peña and Pokutta, Sebastian and Proissl, Manuel and Rebentrost, Patrick and Sahin, Emre and Symons, Benjamin C. B. and Tornow, Sabine and Valls, Víctor and Woerner, Stefan and Wolf-Bauwens, Mira L. and Yard, Jon and Yarkoni, Sheir and Zechiel, Dirk and Zhuk, Sergiy and Zoufal, Christa},
   year={2024},
   month=Oct, pages={718–735}
}

@article{farhi2014quantum,
	title={A Quantum Approximate Optimization Algorithm},
	author={Edward Farhi and Jeffrey Goldstone and Sam Gutmann},
	year={2014},
	eprint={arXiv:1411.4028},
	archivePrefix={arXiv},
	primaryClass={quant-ph},
    journal={arXiv:1411.4028}
}

@article{farhi:qaa,
	title={Quantum Computation by Adiabatic Evolution},
	author={Edward Farhi and Jeffrey Goldstone and Sam Gutmann and Michael Sipser},
	year={2000},
	eprint={arXiv:quant-ph/0001106},
	archivePrefix={arXiv},
	journal=""
}

@misc{farhi2019quantum,
	title={Quantum Supremacy through the Quantum Approximate Optimization Algorithm},
	author={Edward Farhi and Aram W Harrow},
	year={2019},
	eprint={1602.07674},
	archivePrefix={arXiv},
	primaryClass={quant-ph}
}

@article{barren_vqa2021,
	title={On barren plateaus and cost function locality in variational quantum algorithms},
	volume={54},
	ISSN={1751-8121},
	url={http://dx.doi.org/10.1088/1751-8121/abfac7},
	DOI={10.1088/1751-8121/abfac7},
	number={24},
	journal={Journal of Physics A: Mathematical and Theoretical},
	publisher={IOP Publishing},
	author={Uvarov, A V and Biamonte, J D},
	year={2021},
	month={May},
	pages={245301}
}

@article{Cerezo_2021,
	doi = {10.1038/s41467-021-21728-w},
	url = {https://doi.org/10.1038%2Fs41467-021-21728-w},
	year = 2021,
	month = {mar},
	publisher = {Springer Science and Business Media {LLC}},
	volume = {12},
	number = {1},
	author = {M. Cerezo and Akira Sone and Tyler Volkoff and Lukasz Cincio and Patrick J. Coles},
	title = {Cost function dependent barren plateaus in shallow parametrized quantum circuits},
	journal = {Nature Communications}
}

@misc{pt-weighted,
    doi = {10.48550/ARXIV.2201.11785},
    url = {https://arxiv.org/abs/2201.11785},
    author = {Shaydulin, Ruslan and Lotshaw, Phillip C. and Larson, Jeffrey and Ostrowski, James and Humble, Travis S.},
    title = {Parameter Transfer for Quantum Approximate Optimization of Weighted MaxCut},
    publisher = {arXiv},
    year = {2022},
    copyright = {arXiv.org perpetual, non-exclusive license}
}

@article{new-qaoa,
	doi = {10.3390/a12020034},
	url = {https://doi.org/10.3390%2Fa12020034},
	year = 2019,
	month = {feb},
	publisher = {{MDPI} {AG}},
	volume = {12},
	number = {2},
	pages = {34},
	author = {Stuart Hadfield and Zhihui Wang and Bryan O{\textquotesingle}Gorman and Eleanor Rieffel and Davide Venturelli and Rupak Biswas},
	title = {From the Quantum Approximate Optimization Algorithm to a Quantum Alternating Operator Ansatz},
	journal = {Algorithms}
}

@inproceedings{grover-mixer,
	doi = {10.1109/qce49297.2020.00020},
	url = {https://doi.org/10.1109%2Fqce49297.2020.00020},
	year = 2020,
	month = {oct},
	publisher = {{IEEE}},
	author = {Andreas Bartschi and Stephan Eidenbenz},
	title = {Grover Mixers for {QAOA}: Shifting Complexity from Mixer Design to State Preparation},
	booktitle = {2020 {IEEE} International Conference on Quantum Computing and Engineering ({QCE})}
}

@Article{bp-qnn,
    author={McClean, Jarrod R.
    and Boixo, Sergio
    and Smelyanskiy, Vadim N.
    and Babbush, Ryan
    and Neven, Hartmut},
    title={Barren plateaus in quantum neural network training landscapes},
    journal={Nature Communications},
    year={2018},
    month={Nov},
    day={16},
    volume={9},
    number={1},
    pages={4812},
    issn={2041-1723},
    doi={10.1038/s41467-018-07090-4},
    url={https://doi.org/10.1038/s41467-018-07090-4}
}

@Article{hea,
    author={Kandala, Abhinav
    and Mezzacapo, Antonio
    and Temme, Kristan
    and Takita, Maika
    and Brink, Markus
    and Chow, Jerry M.
    and Gambetta, Jay M.},
    title={Hardware-efficient variational quantum eigensolver for small molecules and quantum magnets},
    journal={Nature},
    year={2017},
    month={Sep},
    day={01},
    volume={549},
    number={7671},
    pages={242-246},
    issn={1476-4687},
    doi={10.1038/nature23879},
    url={https://doi.org/10.1038/nature23879}
}

@inproceedings{th-gm-qaoa,
   title={Threshold-Based Quantum Optimization},
   url={http://dx.doi.org/10.1109/QCE52317.2021.00030},
   DOI={10.1109/qce52317.2021.00030},
   booktitle={2021 IEEE International Conference on Quantum Computing and Engineering (QCE)},
   publisher={IEEE},
   author={Golden, John and Bartschi, Andreas and O’Malley, Daniel and Eidenbenz, Stephan},
   year={2021},
   month=oct, pages={137–147}
}

@article{vqe,
   title={The Variational Quantum Eigensolver: A review of methods and best practices},
   volume={986},
   ISSN={0370-1573},
   url={http://dx.doi.org/10.1016/j.physrep.2022.08.003},
   DOI={10.1016/j.physrep.2022.08.003},
   journal={Physics Reports},
   publisher={Elsevier BV},
   author={Tilly, Jules and Chen, Hongxiang and Cao, Shuxiang and Picozzi, Dario and Setia, Kanav and Li, Ying and Grant, Edward and Wossnig, Leonard and Rungger, Ivan and Booth, George H. and Tennyson, Jonathan},
   year={2022},
   month=nov, pages={1–128}
}

@article{constraint-preserving-qaoa,
   title={Constraint Preserving Mixers for the Quantum Approximate Optimization Algorithm},
   volume={15},
   ISSN={1999-4893},
   url={http://dx.doi.org/10.3390/a15060202},
   DOI={10.3390/a15060202},
   number={6},
   journal={Algorithms},
   publisher={MDPI AG},
   author={Fuchs, Franz Georg and Lye, Kjetil Olsen and Møll Nilsen, Halvor and Stasik, Alexander Johannes and Sartor, Giorgio},
   year={2022},
   month=jun, pages={202}
}

@article{if-qaoa,
   title={Penalty-free approach to accelerating constrained quantum optimization},
   volume={112},
   ISSN={2469-9934},
   url={http://dx.doi.org/10.1103/fb5m-cl9m},
   DOI={10.1103/fb5m-cl9m},
   number={6},
   journal={Physical Review A},
   publisher={American Physical Society (APS)},
   author={Bucher, David and Stein, Jonas and Feld, Sebastian and Linnhoff-Popien, Claudia},
   year={2025},
   month=dec
}

@misc{efficient-qaoa,
      title={Efficient QAOA Architecture for Solving Multi-Constrained Optimization Problems}, 
      author={David Bucher and Daniel Porawski and Maximilian Janetschek and Jonas Stein and Corey O'Meara and Giorgio Cortiana and Claudia Linnhoff-Popien},
      year={2025},
      eprint={2506.03115},
      archivePrefix={arXiv},
      primaryClass={quant-ph},
      doi={https://doi.org/10.1109/QCE65121.2025.00048},
      url={https://arxiv.org/abs/2506.03115}, 
}

@article{vqe-vs-qaoa,
   author	 = "Yan, Xinjian and Lee, Xinwei and Cai, Dongsheng and Asai, Nobuyoshi",
   title	 = "Comparison between the performances of General Two-local Ansatzes and QAOA in Max-cut problem",
   journal  = "IPSJ SIG Technical Report",
   volume   = "2024-QS-11",
   number   = "16",
   pages    = "1--8",
   year 	 = "2024",
   month    = mar,
   url      = "https://ipsj.ixsq.nii.ac.jp/ej/?action=pages_view_main&active_action=repository_view_main_item_detail&item_id=233690&item_no=1&page_id=13&block_id=8"
}

@article{bp-review,
    author={Larocca, Mart{\'i}n
    and Thanasilp, Supanut
    and Wang, Samson
    and Sharma, Kunal
    and Biamonte, Jacob
    and Coles, Patrick J.
    and Cincio, Lukasz
    and McClean, Jarrod R.
    and Holmes, Zo{\"e}
    and Cerezo, M.},
    title={Barren plateaus in variational quantum computing},
    journal={Nature Reviews Physics},
    year={2025},
    month={Apr},
    day={01},
    volume={7},
    number={4},
    pages={174-189},
    issn={2522-5820},
    doi={10.1038/s42254-025-00813-9},
    url={https://doi.org/10.1038/s42254-025-00813-9}
}

@article{align-qaoa,
   title={Alignment between initial state and mixer improves QAOA performance for constrained optimization},
   volume={9},
   ISSN={2056-6387},
   url={http://dx.doi.org/10.1038/s41534-023-00787-5},
   DOI={10.1038/s41534-023-00787-5},
   number={1},
   journal={npj Quantum Information},
   publisher={Springer Science and Business Media LLC},
   author={He, Zichang and Shaydulin, Ruslan and Chakrabarti, Shouvanik and Herman, Dylan and Li, Changhao and Sun, Yue and Pistoia, Marco},
   year={2023},
   month=nov
}

@article{toffoli-vqa,
   title={Variational quantum algorithm for constrained combinatorial optimization problems},
   volume={113},
   ISSN={2469-9934},
   url={http://dx.doi.org/10.1103/ykxd-h19w},
   DOI={10.1103/ykxd-h19w},
   number={3},
   journal={Physical Review A},
   publisher={American Physical Society (APS)},
   author={Li, Hui-Min and Han, Yuan-Liang and Wang, Zhi-Xi and Fei, Shao-Ming},
   year={2026},
   month=mar
}

@article{dual-vqe,
   title={Dual variational quantum eigensolver: A quantum algorithm to lower bound the ground-state energy},
   volume={113},
   ISSN={2469-9934},
   url={http://dx.doi.org/10.1103/twrt-y691},
   DOI={10.1103/twrt-y691},
   number={3},
   journal={Physical Review A},
   publisher={American Physical Society (APS)},
   author={Westerheim, Hanna and Chen, Jingxuan and Holmes, Zoë and Luo, Ivy and Nuradha, Theshani and Patel, Dhrumil and Rethinasamy, Soorya and Wang, Kathie and Wilde, Mark M.},
   year={2026},
   month=mar
}

@article{prog-mis,
    doi = {10.1088/1674-1056/addd83},
    url = {https://doi.org/10.1088/1674-1056/addd83},
    year = {2025},
    month = {jul},
    publisher = {Chinese Physical Society and IOP Publishing Ltd},
    volume = {34},
    number = {7},
    pages = {070304},
    author = {Ni, Xiao-Hui and Li, Ling-Xiao and Song, Yan-Qi and Jin, Zheng-Ping and Qin, Su-Juan and Gao, Fei},
    title = {Progressive quantum algorithm for maximum independent set with quantum alternating operator ansatz},
    journal = {Chinese Physics B}
}

@misc{quantinuum-h2-data,
    author = {Quantinuum},
    title = {Quantinuum H2 Product Data Sheet},
    year = {2025},
    organization = {Quantinuum},
    howpublished = "\url{https://docs.quantinuum.com/systems/_static/assets/data_sheets/Quantinuum\%20H2\%20Product\%20Data\%20Sheet.pdf}",
}

@article{f-vqe,
   title={Filtering variational quantum algorithms for combinatorial optimization},
   volume={7},
   ISSN={2058-9565},
   url={http://dx.doi.org/10.1088/2058-9565/ac3e54},
   DOI={10.1088/2058-9565/ac3e54},
   number={1},
   journal={Quantum Science and Technology},
   publisher={IOP Publishing},
   author={Amaro, David and Modica, Carlo and Rosenkranz, Matthias and Fiorentini, Mattia and Benedetti, Marcello and Lubasch, Michael},
   year={2022},
   month=Jan, pages={015021}
}

@misc{qubit-reset,
      title={Qubit-reuse compilation with mid-circuit measurement and reset}, 
      author={Matthew DeCross and Eli Chertkov and Megan Kohagen and Michael Foss-Feig},
      year={2022},
      eprint={2210.08039},
      archivePrefix={arXiv},
      primaryClass={quant-ph},
      url={https://arxiv.org/abs/2210.08039}, 
}

@misc{nv-qaoa,
      title={Non-Variational Quantum Random Access Optimization with Alternating Operator Ansatz}, 
      author={Zichang He and Rudy Raymond and Ruslan Shaydulin and Marco Pistoia},
      year={2025},
      eprint={2502.04277},
      archivePrefix={arXiv},
      primaryClass={quant-ph},
      url={https://arxiv.org/abs/2502.04277}, 
}

@article{gas,
   title={Grover Adaptive Search for Constrained Polynomial Binary Optimization},
   volume={5},
   ISSN={2521-327X},
   url={http://dx.doi.org/10.22331/q-2021-04-08-428},
   DOI={10.22331/q-2021-04-08-428},
   journal={Quantum},
   publisher={Verein zur Forderung des Open Access Publizierens in den Quantenwissenschaften},
   author={Gilliam, Austin and Woerner, Stefan and Gonciulea, Constantin},
   year={2021},
   month=Apr, pages={428}
}

@article{qite-seminal,
   title={Determining eigenstates and thermal states on a quantum computer using quantum imaginary time evolution},
   volume={16},
   ISSN={1745-2481},
   url={http://dx.doi.org/10.1038/s41567-019-0704-4},
   DOI={10.1038/s41567-019-0704-4},
   number={2},
   journal={Nature Physics},
   publisher={Springer Science and Business Media LLC},
   author={Motta, Mario and Sun, Chong and Tan, Adrian T. K. and O’Rourke, Matthew J. and Ye, Erika and Minnich, Austin J. and Brandão, Fernando G. S. L. and Chan, Garnet Kin-Lic},
   year={2019},
   month=nov, pages={205–210}
}

@INPROCEEDINGS{ccop-qite,
  author={Lee, Xin Wei and Lau, Hoong Chuin},
  booktitle={2025 IEEE International Conference on Quantum Computing and Engineering (QCE)}, 
  title={Solving Constrained Combinatorial Optimization Problems with Variational Quantum Imaginary Time Evolution}, 
  year={2025},
  volume={01},
  number={},
  pages={1955-1964},
  doi={10.1109/QCE65121.2025.00213}
}
\end{document}